\documentclass[aps,prc,reprint,superscriptaddress,showpacs,nofootinbib]{revtex4-1}

\usepackage{amsmath,amsfonts,amssymb,amscd,amsthm}
\usepackage{graphicx}
\graphicspath{{./figures/}}
\usepackage{multirow}
\usepackage[usenames,dvipsnames,svgnames,table]{xcolor}
\usepackage{xfrac}
\usepackage{siunitx}
\usepackage{overpic}
\usepackage{bigdelim}
\usepackage{bm}
\usepackage{tabularx}
\usepackage{threeparttable}
\usepackage{ulem}

\newcommand{\nen}{\ensuremath{^{19}\mathrm{Ne}}}
\newcommand{\fln}{\ensuremath{^{19}\mathrm{F}}}
\newcommand{\fle}{\ensuremath{^{18}\mathrm{F}}}
\newcommand{\fpa}{$^{18}$F(p,$\alpha$)$^{15}$O}
\newcommand{\fht}{$^{19}$F($^{3}$He,t)$^{19}$Ne}

\begin{document}


\title{Sub-threshold states in \nen\ relevant to \fpa}


\author{J.~E.~Riley}
\affiliation{Department of Physics, University of York, York, YO10 5DD, United Kingdom}

\author{A.~M.~Laird}
\email[]{alison.laird@york.ac.uk}
\affiliation{Department of Physics, University of York, York, YO10 5DD, United Kingdom}

\author{N.~de~S\'{e}r\'{e}ville}
\email[]{nicolas.de-sereville@ijclab.in2p3.fr}
    \affiliation{Universit\'e Paris-Saclay, CNRS/IN2P3, IJCLab, 91405 Orsay, France}

\author{A.~Parikh}
    \affiliation{Departament de F\'isica, Universitat Polit\`ecnica de Catalunya,
                 E-08036 Barcelona, Spain}
    \affiliation{Institut d'Estudis Espacials de Catalunya (IEEC),
                 E-08034 Barcelona, Spain}
\author{S.~P.~Fox}
    \affiliation{Department of Physics, University of York, York, YO10 5DD, United Kingdom}
\author{F.~Hammache}
    \affiliation{Universit\'e Paris-Saclay, CNRS/IN2P3, IJCLab, 91405 Orsay, France}
\author{I.~Stefan}
    \affiliation{Universit\'e Paris-Saclay, CNRS/IN2P3, IJCLab, 91405 Orsay, France}
\author{P.~Adsley}
    \altaffiliation[Current address: ]{University of the Witwatersrand, South Africa and iThemba LABS, South Africa}
    \affiliation{Universit\'e Paris-Saclay, CNRS/IN2P3, IJCLab, 91405 Orsay, France}
\author{M.~Assi\'e}
    \affiliation{Universit\'e Paris-Saclay, CNRS/IN2P3, IJCLab, 91405 Orsay, France}
\author{B.~Bastin}
    \affiliation{Grand Accélérateur National d'Ions Lourds (GANIL), CEA/DRF-CNRS/IN2P3, 
    Bd. Henri Becquerel, 14076 Caen, France}
\author{F.~Boulay}
    \affiliation{Grand Accélérateur National d'Ions Lourds (GANIL), CEA/DRF-CNRS/IN2P3, 
    Bd. Henri Becquerel, 14076 Caen, France}
\author{A.~Coc}
    \affiliation{Universit\'e Paris-Saclay, CNRS/IN2P3, IJCLab, 91405 Orsay, France}
\author{S.~Franchoo}
    \affiliation{Universit\'e Paris-Saclay, CNRS/IN2P3, IJCLab, 91405 Orsay, France}
\author{R.~Garg}
    \altaffiliation[Current address: ]{School of Physics and Astronomy, University of Edinburgh, Edinburgh, EH9 3FD, United Kingdom}
    \affiliation{Department of Physics, University of York, York, YO10 5DD, United Kingdom}
    
\author{S.~A.~Gillespie}
    \affiliation{Department of Physics, University of York, York, YO10 5DD, United Kingdom}
\author{V.~Guimaraes}
    \affiliation{Universit\'e Paris-Saclay, CNRS/IN2P3, IJCLab, 91405 Orsay, France}
    \affiliation{Instituto de Fisica, Universidade de S\~{a}o Paulo, Rua do Mat\~{a}o, 1371,
S\~{a}o Paulo 05508-090, SP, Brazil}
\author{C.~Hamadache}
    \affiliation{Universit\'e Paris-Saclay, CNRS/IN2P3, IJCLab, 91405 Orsay, France}
\author{N.~Hubbard}
    \affiliation{Department of Physics, University of York, York, YO10 5DD, United Kingdom}
\author{J.~Kiener}
    \affiliation{Universit\'e Paris-Saclay, CNRS/IN2P3, IJCLab, 91405 Orsay, France}
\author{A.~Lefebvre-Schuhl}
    \affiliation{Universit\'e Paris-Saclay, CNRS/IN2P3, IJCLab, 91405 Orsay, France}
\author{F.~de~Oliveira Santos}
    \affiliation{Grand Accélérateur National d'Ions Lourds (GANIL), CEA/DRF-CNRS/IN2P3, 
    Bd. Henri Becquerel, 14076 Caen, France}
\author{A.~Remadi}
    \affiliation{Grand Accélérateur National d'Ions Lourds (GANIL), CEA/DRF-CNRS/IN2P3, 
    Bd. Henri Becquerel, 14076 Caen, France}
\author{L.~Perrot}
    \affiliation{Universit\'e Paris-Saclay, CNRS/IN2P3, IJCLab, 91405 Orsay, France}
\author{D.~Suzuki}
    \affiliation{Universit\'e Paris-Saclay, CNRS/IN2P3, IJCLab, 91405 Orsay, France}
\author{G.~Verde}
    \affiliation{Universit\'e Paris-Saclay, CNRS/IN2P3, IJCLab, 91405 Orsay, France}
\author{V.~Tatischeff}
    \affiliation{Universit\'e Paris-Saclay, CNRS/IN2P3, IJCLab, 91405 Orsay, France}
\author{M.~Williams}
    \affiliation{Department of Physics, University of York, York, YO10 5DD, United Kingdom}

\date{\today}

\begin{abstract}
   \begin{description}
      \item[Background] Classical novae result from thermonuclear explosions producing several $\gamma$-ray emitters which are prime targets for satellites observing in the MeV range. The early $\leq511$~keV gamma-ray emission depends critically on the \fpa\ reaction rate which, despite many experimental and theoretical efforts, still remains uncertain.
      \item[Purpose] One of the main uncertainties in the \fpa\ reaction rate is the contribution in the Gamow window of interference between sub-threshold \nen\ states and known broad states at higher energies. Therefore the goal of this work is to clarify the existence and the nature of these sub-threshold states.
      \item[Method] States in the \nen\ compound nucleus were studied at the Tandem-ALTO facility using the \fht\ charge exchange reaction. Tritons were detected with an Enge Split-pole spectrometer while decaying protons or $\alpha$-particles from unbound \nen\ states were collected, in coincidence, with a double-sided silicon strip detector array. Angular correlations were extracted and constraints on the spin and parity of decaying states established.
      \item[Results] The coincidence yield at E$_x$ = 6.29 MeV was observed to be high spin, supporting the conclusion that it is indeed a doublet consisting of high spin and low spin components. Evidence for a broad, low spin state was observed around 6 MeV. Branching ratios were extracted for several states above the proton threshold and were found to be consistent with the literature. $\mathcal{R}$-matrix calculations show the relative contribution of sub-threshold states to the astrophysically important energy region above the proton threshold.
      \item[Conclusions] The levels schemes of \nen\ and $^{19}$F are still not sufficiently well known and further studies of the analogue assignments are needed. The tentative broad state at 6 MeV may only play a role if the reduced proton width is large. 
   \end{description}
\end{abstract}

\maketitle


\section{Introduction} \label{sec:Intro}
Classical novae outbursts are phenomena taking place in a binary system made up of a white dwarf accreting material from its companion star~\cite{Bod08}. This material is progressively heated and compressed at the surface of the white dwarf until it reaches the ignition temperatures for hydrogen burning in degenerate conditions. During this explosive burning, nucleosynthesis takes place in a fully convective envelope and the newly synthesized material is ejected into the circumstellar medium. 


The most intense $\gamma$-ray line emission from classical novae is predicted to come from the $\beta^+$-decay of \fle\, producing a signature at and below 511~keV from positron annihilation. Due to the short \fle\ half-life ($T_{1/2}=110$~min) similar to the transparency time of the ejected envelope to $\gamma$-rays, the $\gamma$-ray emission at $\leq511$~keV would give unique insights into the details of the expanding shell (velocity, material profile). Precise knowledge of the yield of \fle\ produced during the explosion is therefore crucial for interpreting future observations. After several decades of experimental and theoretical work the main remaining nuclear physics uncertainty affecting model predictions of the  \fle\ yield arises from the \fpa\ reaction rate. 

Due to its importance, considerable experimental effort has been expended in studying the \fpa\ reaction. The astrophysically-relevant energy range covers 50 -- 350 keV in the center of mass, and a number of direct measurements have been performed down to energies of 250 keV (see~\cite{Bardayan2001, Bardayan2002, Chae2006, DeSereville2009, Beer2011}). Measurements at lower energies, however, are limited by the currently available $^{18}$F beam intensities. 

In the absence of direct measurements across the full energy range, a variety of indirect techniques have been exploited to determine the \nen~level information necessary to allow the \fpa\ reaction rate to be calculated. Resonant elastic scattering studies were performed by Bardayan et al.~\cite{Bardayan2004}, Murphy et al.~\cite{Murphy2009} and Mountford et al.~\cite{Mountford2012} and parameters for several resonances at and above $E_r^{c.m.}$ = 665 keV were determined. Transfer reaction studies have also been performed, including $^{18}$F(d,p)$^{19}$F~\cite{Kozub2006,DeSereville2003}, $^{18}$F(d,n)\nen~\cite{Adekola2011} and $^{20}$Ne(p,d)\nen~\cite{Bardayan2015} to explore the region close to the proton threshold. A detailed summary of the known level information will be presented in a future paper~\cite{chetec}, but here we summarise the situation relevant for the present work.

Due to the presence of broad states at $E_r^{c.m.}$ = 665 keV (3/2$^+$) and around 1468 keV (1/2$^+$), 1/2$^+$ and 3/2$^+$ states close to the proton threshold at 6.411 MeV can have a significant impact on the $S$-factor, despite being narrow. The interference between these states and the tails of the broad states can significantly change the predicted $S$-factor in the energy region between 50 and 300 keV (between 300 and 350 keV the $E_r^{c.m.}$ = 331 keV, $J^\pi=3/2^-$ resonance dominates). Therefore it is critical to constrain the location of these states. From the mirror nucleus, $^{19}$F, level scheme two 3/2$^+$ states and one 1/2$^+$ state are expected in the region around the proton threshold ($E_x =$ 6.0 -- 6.6 MeV).

Just below the p+$^{18}$F threshold, an $\ell=0$ state was observed at around 6.290 MeV by Adekola et al. using the $^{18}$F(d,n)$^{19}$Ne reaction~\cite{Adekola2012}. The state was subsequently identified as a 1/2$^+$ state by Bardayan et al. through the $^{20}$Ne(p,d)$^{19}$Ne reaction~\cite{Bardayan2015}. More recently Laird et al.~\cite{Laird2013} and Parikh et al.~\cite{Parikh2015} performed high resolution $^{19}$F($^3$He,t)\nen\ measurements of states above 6 MeV, particularly focusing on the near proton threshold region. These works found the state around 6.290 MeV to be inconsistent with a single low-spin assignment and suggested a doublet with at least one high spin component. 

Kahl et al.~\cite{Kahl19} also used the $^{19}$F($^3$He,t)\nen\ reaction to populate \nen\ states but at intermediate energies with triton detection at very forward angles, to specifically identify $\Delta$L = 0 transitions indicating 1/2$^+$ and 3/2$^+$ states. A state at 6.13 MeV was thus observed, agreeing with one of the two possible spin assignments suggested by Laird et al.. A possible $\Delta$L = 0 contribution to the region around 6.289 MeV was also reported. Based on the required energy shift from the possible analogue states in $^{19}$F, Kahl et al. concluded that the 6.13 MeV is most likely 1/2$^+$, rather than 3/2$^+$, and the component around 6.289 MeV is therefore 3/2$^+$. This is in contradiction to the findings of Bardayan et al.. Indeed the angular distribution presented by Bardayan et al. (Fig 2 in that work) from the $^{20}$Ne(p,d)\nen~reaction is inconsistent with a 3/2$^+$ assignment. It should also be noted that the neutron threshold in 
$^{19}$F is at 10.432 MeV, around 4 MeV higher than the proton threshold in $^{19}$Ne. The Thomas-Ehrman shift on such states could therefore be substantial.

Visser et al.~\cite{Visser2004} studied the proton and $\alpha$-particle decay of excited states in \nen\ populated via the $^{19}$F($^3$He,t)\nen\ reaction. Here, a 25 MeV $^3$He beam was incident on a CaF$_2$ target and the resulting tritons were detected at the focal plane of the Yale Enge magnetic spectrometer. Coincident decay particles were detected in an array of single sided silicon detectors. Decay probability distributions were extracted and \nen~level parameters deduced (see Section IV.B for further details). However, no results for the region just below the proton threshold were reported.

Finally, Hall et al.~\cite{Hall19} also used the $^{19}$F($^3$He,t)\nen\ reaction to populate \nen\ states and then studied the $\gamma$-decays from the de-excitation of these states. A 30 MeV $^3$He beam was incident on a CaF$_2$ target and the resulting tritons were detected with the ORRUBA silicon array. Coincident $\gamma$-rays were detected in the Compton-suppressed high-purity germanium detector array Gammasphere. Triton-$\gamma$-$\gamma$ coincidences suggested two 3/2$^+$ states at 13 and 31 keV above the proton threshold. These assignments suggests either disagreement with the assignments of Laird et al. for this energy region or the presence of additional states not resolved in previous studies. Furthermore, as only two 3/2$^+$ states are known in $^{19}$F in this energy region, either the Hall et al. and Kahl et al. assignments are in contradiction, or there are unobserved $\ell=0$ states present in $^{19}$F.

Hall et al. also observed decays from the sub-threshold state at 6.292 MeV and suggested an 11/2$^+$ assignment based on the similarity to the decay scheme of the 6.500 MeV in $^{19}$F. This assignment indicates an energy shift of 208 keV. Such a large energy shift suggests, therefore, that the average shift of 50 $\pm$ 30 keV estimated by Nesaraja et al. should be considered with caution.    

It is clear, therefore, that the location, and indeed number, of 3/2$^+$ and 1/2$^+$ is still uncertain. The situation is further confused by the prediction of a broad 1/2$^+$ around 6 MeV by Dufour and Descouvement~\cite{Dufour2007} which is as yet unobserved.

Here we report on a study of the $^{19}$Ne level scheme for excitation energies between 6- and 7.5-MeV. The $^{19}$F($^3$He,t)\nen\ reaction was used to populate and identify the relevant states. From the coincident detection of $\alpha$-particles from the decay of $^{19}$Ne, information on the spin-parity and branching ratios was extracted, which did not depend on the charge exchange reaction model assumed. Section II below describes the experimental setup and technique used, and Section III details the data analysis methodology for singles and then coincidence events. The results are presented in Section IV and the interpretation given in Section V. We summarise and conclude in Section VI.

\section{Experiment} \label{sec:Exp}
The \fht\ reaction was studied using the tandem accelerator at the ALTO facility in Orsay, France. A beam of $^{3}$He was produced by the duoplasmatron ion source, accelerated to 25 MeV and transported to the object focal point of an Enge Split-pole magnetic spectrometer~\cite{Spencer1967} with a typical intensity of 70~enA where it impinged upon the targets. Two CaF$_2$ targets with a thickness of 100~$\mu$g/cm$^{2}$ and 200~$\mu$g/cm$^{2}$ were used during the course of the experiment, both backed onto a foil of \textsuperscript{nat}C. Light reaction ejectiles entered the Split-pole spectrometer positioned 10$^{\circ}$ from the beam line through a rectangular aperture. Though the nominal aperture covers 1.7~msr, it was opened to an extent covering 3.3~msr to maximise the triton yield, and corresponding to an angular acceptance of $\pm3^\circ$. However the presence of optical aberrations in these conditions degraded the energy resolution, which was measured to be $\approx$85~keV (FWHM). Moreover, due to the horizontal asymmetry of the aperture, the effective detection angle was 12$^{\circ}$. Light reaction particles were momentum analyzed and focused on the focal-plane detection system~\cite{Mar75}, consisting of a position sensitive gas chamber (where the position ($Pos$) and anode wire signal ($Wire$) are recorded), a $\Delta E$ proportional gas counter and a plastic scintillator to measure the remaining energy. 

\begin{figure}[htpb]
    \begin{overpic}[width=\columnwidth]{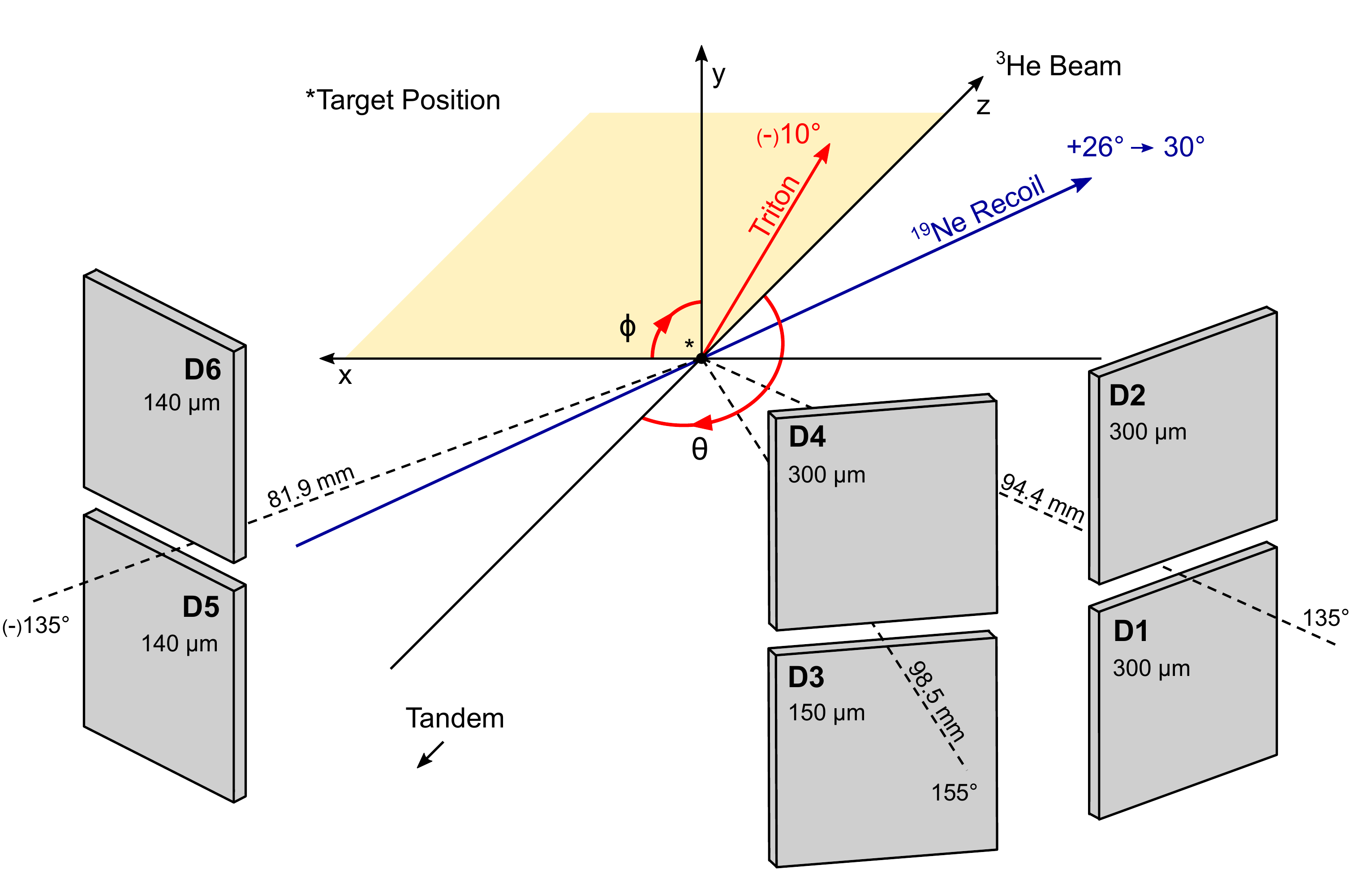}
	\tiny
	\put(50.5,31.7){$\bm{\theta}$}
	\put(46.5,42.5){$\bm{\phi}$}
    \end{overpic}
    \caption{Schematic diagram of the reaction chamber in front of the Split-pole. The z axis points along the beam. The $^{19}$Ne recoil direction causes a rotation to the reference frame of $\approx28^{\circ}$ from which $\theta_{\mathrm{c.m.}}$ is calculated.}
    \label{fig:Geometry}
\end{figure}

In addition to measuring tritons, $\alpha$-particle and proton decays from unbound $^{19}$Ne states were detected in a silicon array placed around the target at backward angles in the laboratory frame (see Fig.~\ref{fig:Geometry}). Six double sided silicon strip detectors (DSSSDs) were mounted on three independent mechanical supports, each holding a pair of detectors, one positioned above the other. Each DSSSD is a square detector of $5\times5$~cm$^2$ with 16 strips on each side (W1 models from Micron Semiconductor Ltd) with thicknesses of 140~$\mu$m or 300~$\mu$m. The mechanical mounts were located upstream of the target at 113$^{\circ}$ (D1 and D2), $^{\tiny(-)}$135$^{\circ}$ (D5 and D6) and 155$^{\circ}$ (D3 and D4) in the laboratory frame, providing an angular range of 91$^{\circ}$ (5$^{\circ}$ at the extreme were obscured by the target mount) and a total solid angle of $\Omega$ = 1.44~sr. A 1-cm thick steel shield was placed in a vertical median plane defined by the target ladder so that the DSSSD array was shielded from the activation of the 0$^{\circ}$ Faraday cup inside the reaction chamber. Energy calibration of the DSSSD array was undertaken using a triple $\alpha$-source placed at the target position.

Since the Split-pole is positioned at an angle of 10$^{\circ}$, the \nen\ recoil direction is between  26$^\circ$ and 30$^\circ$ (see Fig.~\ref{fig:Geometry}) depending on the excited state considered within the Spit-Pole acceptance. The determination of the center of mass angle ($\theta_{c.m.}$) of the decay particles detected in the DSSSD array should then account for the \nen\ recoil direction. In the present case $\theta_{c.m.}$ covers a range between 90$^\circ$ and 172$^\circ$ covering the full possible decay range angle. Such a wide angular coverage allows for a complete measurement of the angular correlation and a reliable determination of the branching ratios.

The Split-pole plastic scintillator was used to trigger the data acquisition system. Information from the Split-pole focal plane detectors and DSSSDs were recorded along with timing information, relative to the event trigger, from the W1 detector front strips. To compensate for the flight time of the tritons, the signals from the DSSSDs were stretched using the shaping time of the Mesytec STM-16+ shaping amplifiers, such that triton recoil and $^{19}$Ne* decay products corresponding to the same event appeared within the 2 $\mu$s DAQ timing window.

\section{Data Analysis} \label{sec:Analysis}

\subsection{Singles Events} \label{sec:singles}
\begin{figure}[!htpb]
    \includegraphics[width=\columnwidth]{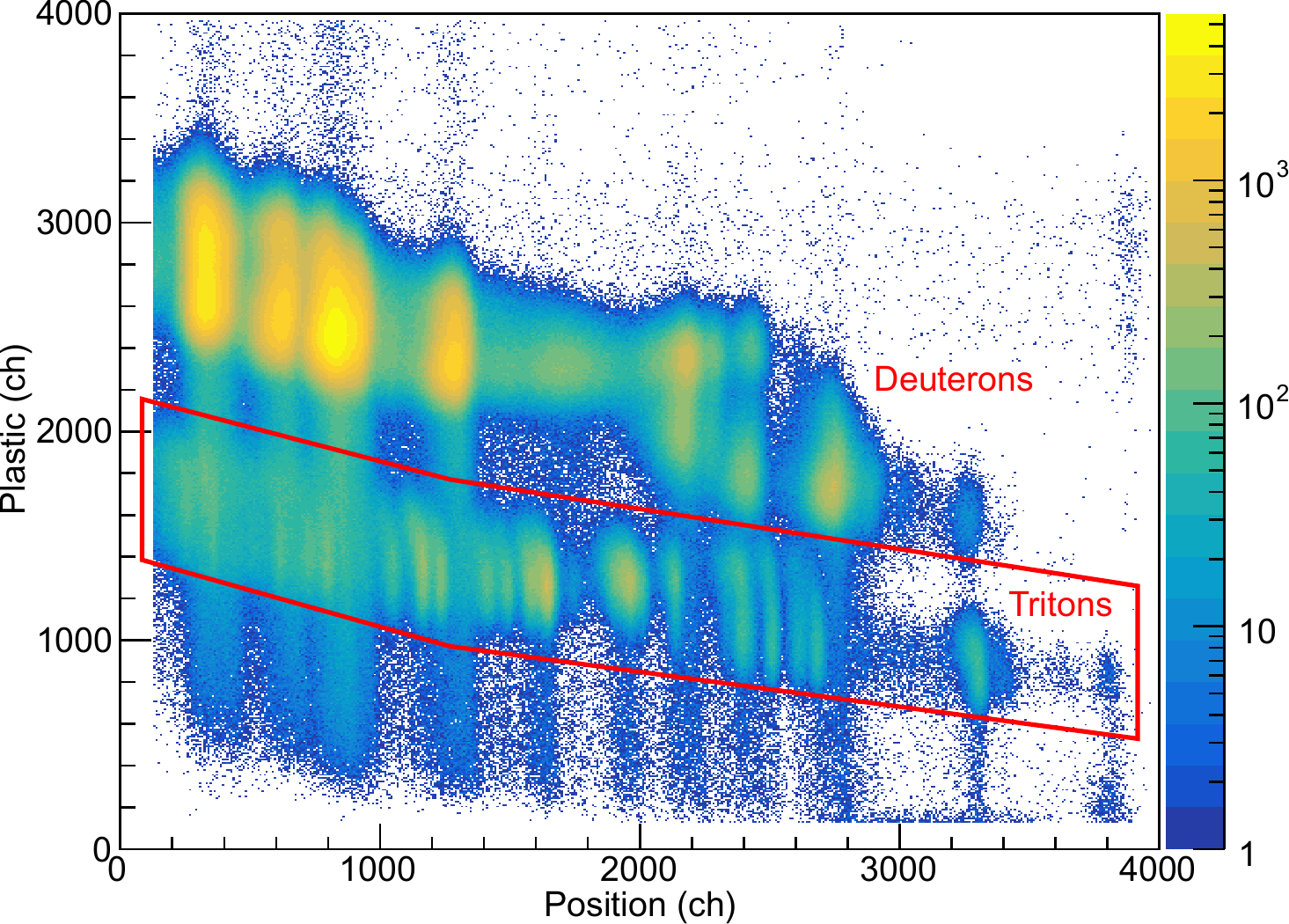}
    \caption{(Colour online) Split-pole Plastic against Position signals to separate particles at the focal plane by Z and A. The triton selection cut is highlighted in red. Background from partial energy deposition of deuterons ‘bleeds’ into the triton region.}
    \label{fig:PID}
\end{figure}

Several combinations of measured quantities in the focal-plane detectors (residual energy v.s. $Wire$, residual energy v.s. $Pos$, $Pos$ v.s. $Wire$ and $Pos$ v.s. $\Delta E$) were used to identify the tritons from deuterons. Fig.~\ref{fig:PID} shows the residual energy versus position measurement. Given the $Q$-values of ($^3$He,t) reactions on possible target contaminants ($^{nat}$C, $^{16}$O and $^{nat}$Ca) and the magnetic field considered for the present measurement (1.42~T) only tritons from the \fht\ reaction reached the focal plane.

\begin{figure*}[htpb]
    \includegraphics[width=\textwidth]{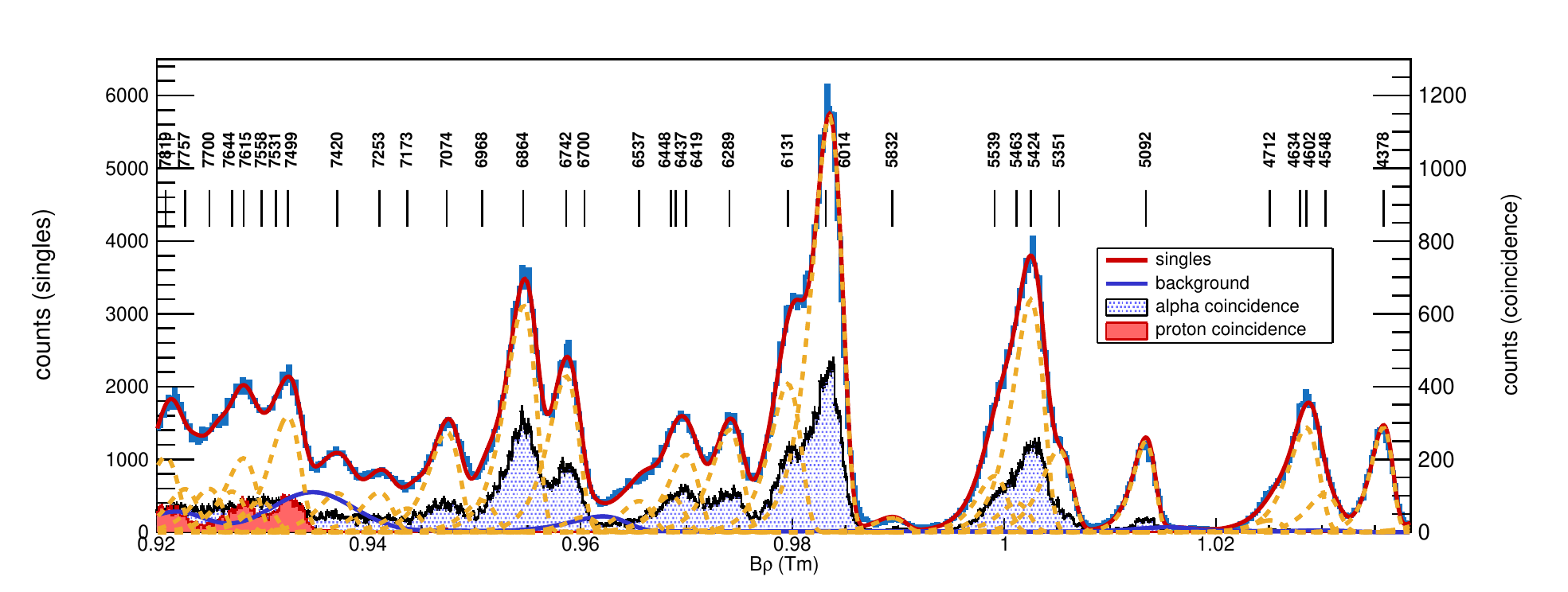}
    \caption{(Colour online) Split-pole focal plane showing the gated triton spectrum. Also shown are events with $\alpha$-particle detection (blue) and proton detection (red). The result of the fit function has been shown in red with individual states in yellow. The only source of background in the focal plane originating from ($^{3}$He,$d$) has been shown in blue.}
    \label{fig:FP}
\end{figure*}

Once the tritons were identified and selected in the focal-plane detector data, their position spectrum was obtained (see Fig.~\ref{fig:FP}). Six well-known isolated and well populated \nen\ states [4379.1~(22), 5092~(6), 6013~(7), 6742~(2), 6864~(2) and 7076~(2)~keV] across the whole focal plane were used to calibrate the focal-plane position detector. A relation between the radius of curvature $\rho$ and the focal-plane position was extracted, and well described with a one-degree polynomial function. 

A functional including peak and background components was constructed to describe the triton magnetic rigidity spectrum and the best fit was obtained after a least-squares minimization procedure (see Fig.~\ref{fig:FP}, red curve). Owing to the extended spectrometer acceptance used in the present experiment the triton peaks exhibit a low energy tail, and the line shape was therefore described by a skewed normal distribution. In most cases the analyzed \nen\ states have a natural width smaller than the experimental resolution, and a common width was therefore used as a free parameter in the fitting procedure. In the case of \nen\ states having a natural width larger than the resolution the width of the skewed normal distribution was set to the natural width, with an allowed variation during the fitting process equal to its documented uncertainty. The centroid of the peak was allowed to vary within the uncertainty associated to the corresponding \nen\ state energy. 

The main source of background in the triton magnetic rigidity spectrum comes from deuterons produced by the ($^3$He,d) reactions. Indeed deuteron events were observed to bleed into the triton selection cut, as can be seen in Fig.~\ref{fig:PID}. Deuteron magnetic rigidity spectra obtained by gating on events immediately above and below the triton locus showed the same shape. This shape was then used as a template for the deuteron background in the fitting procedure with its amplitude as a unique free parameter. The background contribution obtained in the best fit (see Fig.~\ref{fig:FP}, purple line) was found in good agreement with that  expected, based on the amplitude of the deuteron magnetic rigidity spectra aforementioned.

The states included in the fit were taken from references \cite{Nesaraja2007, Laird2013, Parikh2015, Murphy2009, Dalouzy2009, Adekola2012}. 
The best fit of the triton magnetic rigidity spectrum (see Fig.~\ref{fig:FP}; red line) provides a very good description of the data, with the exception of the region at the lower excitation energy side of the \nen\ state at 6014~keV, and the region at slightly lower excitation energies than the 6742 keV state. The former region will be discussed in further detail in Sec.~\ref{sec:Broad}. For the latter, the fit is improved by the inclusion of two additional states. Nesaraja et al. \cite{Nesaraja2007} does indeed predict two states in this region, at 6504 and 6542 keV, one of which is consistent with that found by Cherubini et al. \cite{Cherubini2015}.

\subsection{Coincident Events}
\begin{figure}[htpb]
    \includegraphics[width=\columnwidth]{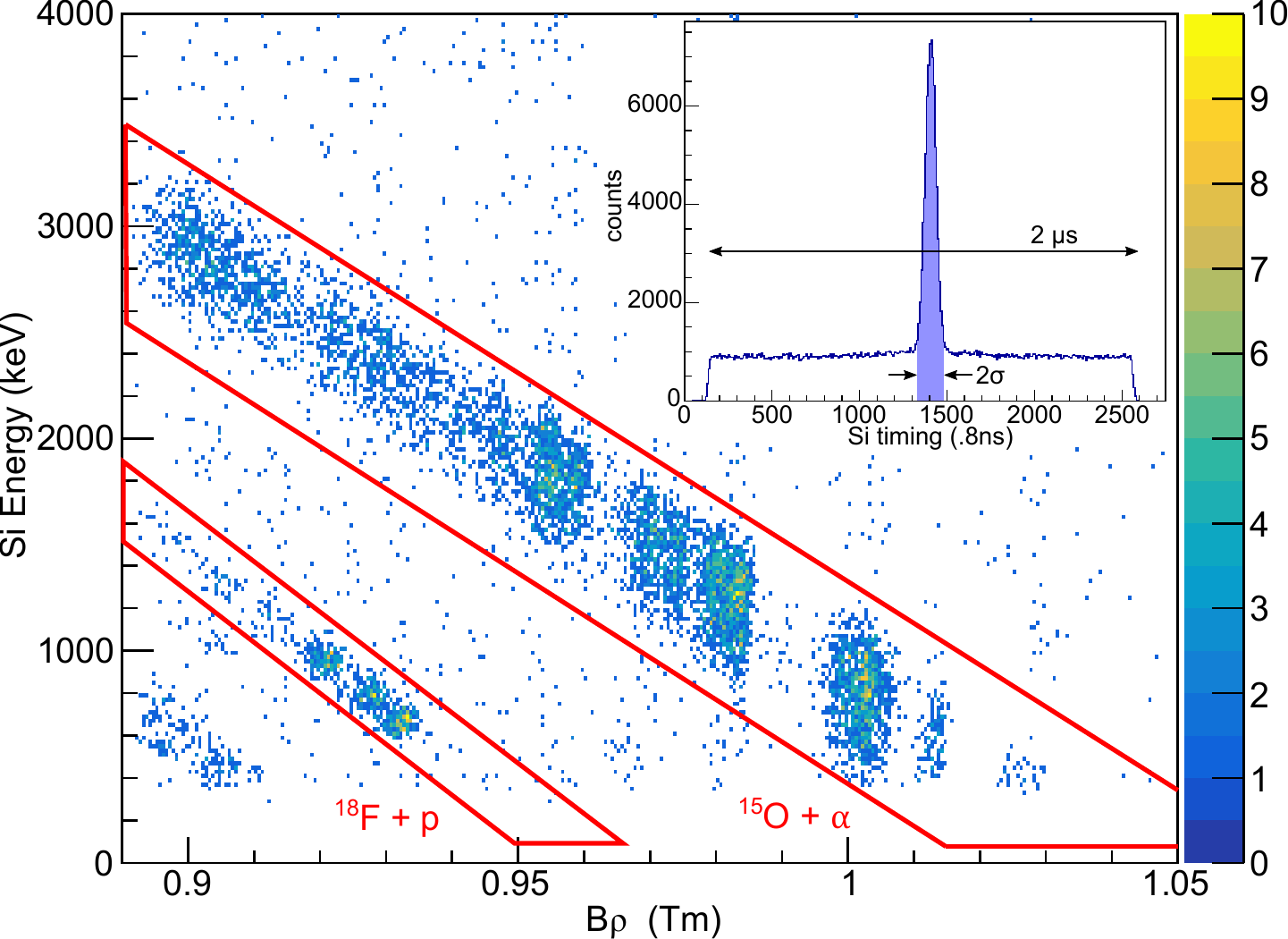}
    \caption{(Color online) Energy deposited in the silicon array against Split-pole position (energy) after applying a time of flight (ToF) cut seen in the inset. The two loci identified in red were used to select proton and $\alpha$-particles coincident event type. The third locus at the bottom left corner corresponds to events populating the first excited state of $^{19}$Ne and were not analysed.}
    \label{fig:EvE}
\end{figure}

Particle decays coincident with triton detection were selected on the basis of timing, whereby true coincidences were identified by a prominent peak above a background of unrelated decays within the reaction chamber as can be observed in the inset of Fig.~\ref{fig:EvE}. Valid decaying events in the DSSSD array were additionally selected when a similar energy deposit (within 2-$\sigma$) between the $p$- and $n$-side of the semiconductor was recorded. The energy deposited in the DSSSD array for coincident events, fulfilling the previous two conditions, as a function of the corresponding triton magnetic rigidity is shown in Fig.~\ref{fig:EvE}. Two kinematic loci with different slopes are observed corresponding to coincident $\alpha$-particle decays to the ground state of $^{15}$O (J$^{\pi}$ = 1/2$^{-}$) and coincident proton decays to the ground state of $^{18}$F (J$^{\pi}$ = 1/2$^{+}$). Software gates associated to each type of events are represented in red.

In order to correctly extract the angular distribution of the $^{19}$Ne* decay particles, the geometry of the DSSSD array was rotated and boosted into centre of mass (c.m.) coordinates. As reaction kinematics are dependant on the \nen\ state populated, this procedure was performed separately for each state of interest since the associated \nen\ recoil direction is changing. These data were then separated into a number of distinct but equal angular ranges, the number of which depended on the population of the state. Focal plane spectra of tritons with a confirmed $\alpha$-particle or proton coincidence were plotted for each angular bin and fitted with the same function used for the triton singles, allowing only peak normalization to vary. 

The yield of each angular bin was extracted by integrating the function used to fit the state. The background of coincidences (seen in the inset of Fig.~\ref{fig:EvE}) was then subtracted proportionally from each angular yield. Finally the geometrical efficiencies for each bin were calculated by performing \textsc{Geant}4~\cite{Agostinelli2003} simulations of the experiment (assuming an isotropic distribution of $^{19}$Ne* decays) constructed using the NPTool package~\cite{Matta2016}.

High energy thresholds in the DSSSD array along with the lower decay probability from astrophysically relevant states meant that resonance parameters could not be extracted with confidence from the proton coincidence data for states below 7500 keV. Alpha-particle decay data were sufficient, however, for the analysis of $^{19}$Ne states both above and below S$_{p}$ relevant to the \fpa\ reaction. The angular probability distributions of the emitted $\alpha$-particles for the 6289, 6742, 6864, 7076 and 7500 keV states in $^{19}$Ne have been extracted and are discussed in Section~\ref{sec:AngCor}. It was not possible to extract these distributions for any other states due to the resolution of the focal plane data.

\section{Results} \label{sec:Results}
\subsection{Evidence for a broad state at $E_x$ = 6~MeV} \label{sec:Broad}
\begin{figure}[htpb]
    \includegraphics[width=8cm]{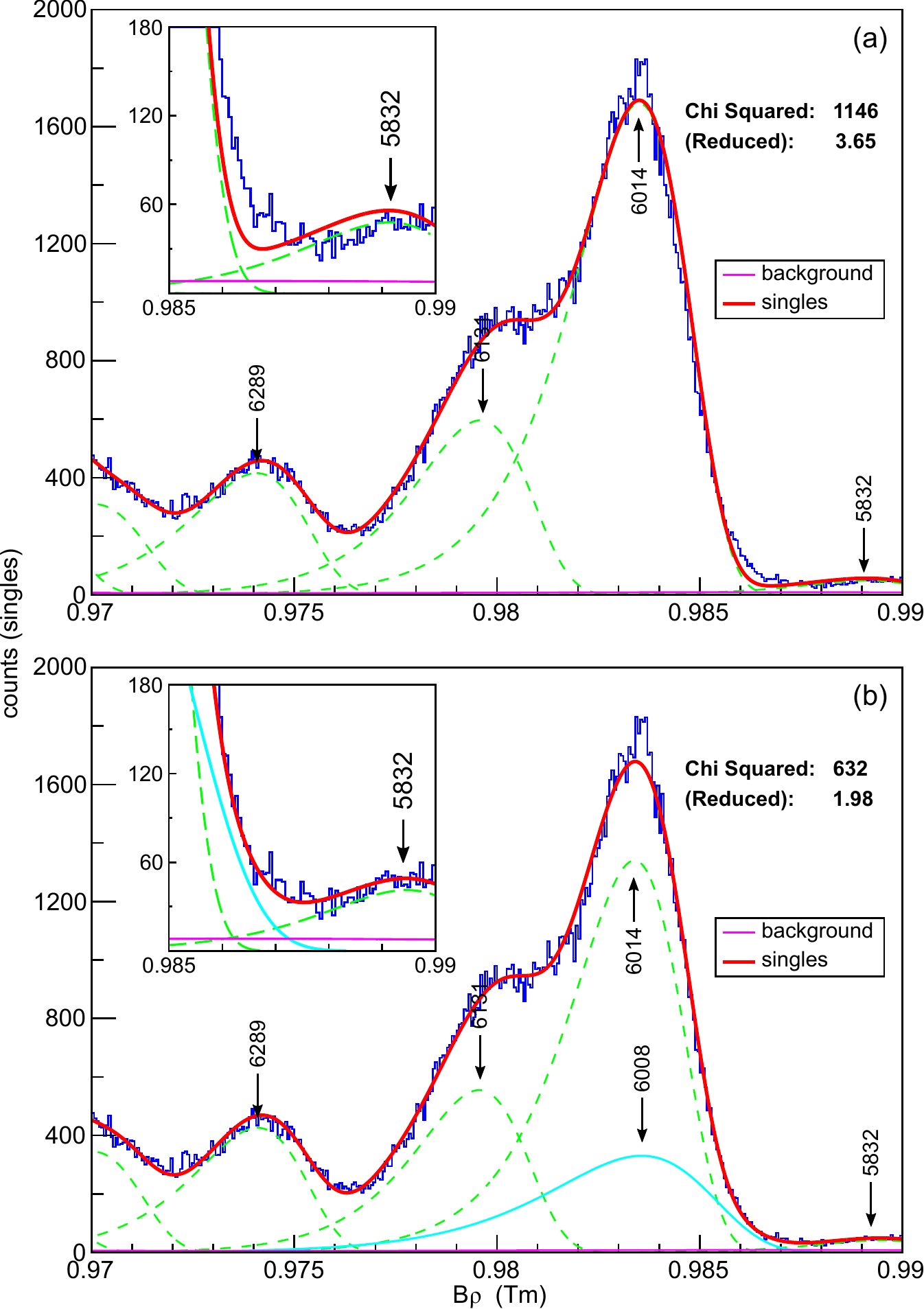}
    \caption{(Color online) Fitted Split-pole focal plane spectra focussed on the $-$400 keV sub-threshold region. The inclusion of an additional broad state to the fit function in panel \textbf{b} shows a marked improvement to the goodness of fit.}
    \label{fig:BroadFit}
\end{figure}

The triton magnetic rigidity spectrum is described extremely well by the best fit, performed as detailed in Section~\ref{sec:singles}, with the exception of the region on the lower excitation energy side of the 6014 keV state. Here, the best fit underestimates the low energy (high magnetic rigidity) side of the triton peak and a significant excess of counts can be observed at a magnetic rigidity of 0.987~Tm as shown in the upper panel of Fig.~\ref{fig:BroadFit}. This region is, however, well described if an additional state, characterized by three independent parameters associated to a skewed normal distribution, is included in the minimization procedure. Results are shown in Fig.~\ref{fig:BroadFit} (lower panel) and the goodness of fit is largely improved with a reduced chi-squared $\chi^2$/ndf~=~1.98 instead of 3.65. The improved fit therefore suggests the presence of an additional state in \nen\ at an energy of 6008~(20)~keV with a total width of $\Gamma = 124$~(25)~keV. A similar analysis of the total coincident spectrum also indicates the presence of a state with compatible energy and total width.

Although the evidence for this additional state in the present work is tentative, further investigation into the properties of this state is justified to constrain its possible impact on the \fpa\ cross section.

It should be noted that the present work was performed 
at the same incident beam energy as that of Laird et al.~\cite{Laird2013}
but with a larger angular acceptance, centred at a slightly different angle. A comparison can be made with Fig. 1 in that work, bearing in mind the different detection angle and therefore different relative population of states.
Although the energy resolution is significantly better and the known states are well separated, the limited statistics prevent any conclusion regarding the existence of the broad state.

The events, neither in singles nor coincidences, corresponding to this possible new state cannot be unambiguously separated from those of the 6014, 6072, and 6100 keV states, and so an angular correlation could not be reliably extracted. Therefore, no constraint can be deduced on the spin-parity of this state from such an approach. However comparison of the extracted width of this state with the Wigner limit does provide stringent constraints. Since the 6008~keV state is below the p+\fle\ threshold, its total width is equal to its $\alpha$-particle partial width. The Wigner limit, defined as $\Gamma^{Wigner}_\alpha = 3\hbar^2/(\mu r^2) \times P_\ell(r,E)$ where $\mu$ is the reduced mass of the $\alpha$+$^{15}$O channel and $P_\ell(r,E)$ is the penetrability for the Coulomb and centrifugal barriers, was calculated for various transferred angular momentum $\ell$. The experimental width ($\Gamma=124$~keV) exceeds the Wigner limit for $\ell\geq2$ giving strong constraints on the spin and parity of this broad state. In the $\alpha$+$^{15}$O channel, $\ell=0$ corresponds to $J^\pi=1/2^-$ and $\ell=1$ corresponds to $J^\pi=1/2^+$ or 3/2$^+$. The Wigner limits are $\Gamma^{Wigner}_\alpha(\ell=0)=265$~keV and $\Gamma^{Wigner}_\alpha(\ell=1)=165$~keV. In either case, the broad state at 6008~keV would therefore be a strong $\alpha$-cluster state since its dimensionless $\alpha$-particle reduced width $\theta^2_\alpha=\Gamma/\Gamma_\alpha^{Wigner}$ is greater than 50\%.

\subsection{Angular correlations} \label{sec:AngCor}
\subsubsection{Formalism and method}
The angular variation in decay product emission is governed by the orbital  angular momentum $l$ transferred to the decay particle and the spin $J$ of the originating state. Particle decay distribution from isolated nuclear levels are described by a summation of even terms of the Legendre polynomials, $P_{k}(\cos(\theta))$, truncated at a maximum value $k_{max} = min(2l,2J)$. Following the formalism derived by Pronko \& Lindgren~\cite{Pronko1972}, the correlation function $W(\theta)$ in its most general form reads
\begin{eqnarray}
    W(\theta) = \sum_{mll'skr} P(m)\: A(Jll'smk)\: (2-\delta_{ll'}) \nonumber \\
    \times\: X^{r}(ll')\: Y(s)\: Q(k)\: P_{k}(\cos(\theta)).
    \label{eqn:AngDist}
\end{eqnarray}
The population of each ($2J+1)$ magnetic substate $m$ is given by $P(m)$, while the population of each exit channel spin $s$ is given by $Y(s)$. The orbital angular momenta $l$ and $l'=l+2$ of the decaying particle represent the different possible values when several exit channel spins are allowed. The interference from the competing orbital angular momenta is accounted by the mixing ratio $X^{r}(ll')$. The term $A(Jll'smk)$ is a product of Clebsch-Gordon and Racah coefficients which is evaluated numerically, and the solid angle correction $Q(k)$ is equal to 1 given the very small detection angle of the decaying particles subtended by a pixel of the DSSSD array~\cite{Pronko1972}.

In case of $\alpha$-particle (0$^{+}$) decays from \nen\ excited states to $^{15}$O ground state (1/2$^{-}$) a single channel spin $s=1/2$ is allowed, and only a single orbital angular momentum $l$ is possible for a given spin $J$ of the \nen\ decaying state. The general form of Eq.~\ref{eqn:AngDist} then simplifies to 
\begin{equation}
    W_{\alpha}(\theta) = \sum_{m,k} P(m)\: A(Jll\tfrac{1}{2}mk)\: P_{k}(\cos(\theta)).
    \label{eqn:AngDistSimp}
\end{equation}
Owing to the properties of the $A(Jll\tfrac{1}{2}mk)$ term in case of a channel spin $s=1/2$, it evaluates to the same value for decaying states having the same spin independently of their parity~\cite{Pronko1972}. It is also worth noting that the $A(Jll'smk$) term gives identical results if $m$ is replaced by $-m$. Therefore, experimental $\alpha$-particle angular correlations were fitted using Eq.~\ref{eqn:AngDistSimp} where the sum of the population of magnetic substates $P(m)$$+P(-m)$ were considered as free parameters. In addition the magnetic substate population should fulfill the two following relations:  $0\leq P(m)+P(-m)\leq1$ and $\sum_{m}P(m)=1$. For a given \nen\ state there are $(2J-1)/2$ free parameters, and an additional overall scaling factor (not present in Eq.~\ref{eqn:AngDistSimp}).

The case of proton (1/2$^+$) emission from \nen\ excited states to \fle\ ground state (1$^+$) is more complicated since two channel spin $s=1/2$ and $s=3/2$ are possible. In this case the general angular correlation function given by Eq.~\ref{eqn:AngDist} is used. The fitting of the experimental proton angular distribution is then performed as for the $\alpha$-decay case with one additional free parameter $Y(1/2)$ and the condition $\sum_{s}Y(s)=1$. The minimum value of the orbital angular momentum which couples the proton to the \nen\ decaying state is chosen, which sets the mixing term $X^r(ll')$ equal to one.

\subsubsection{Results}
\begin{figure}[htpb]
    \includegraphics[width=9cm]{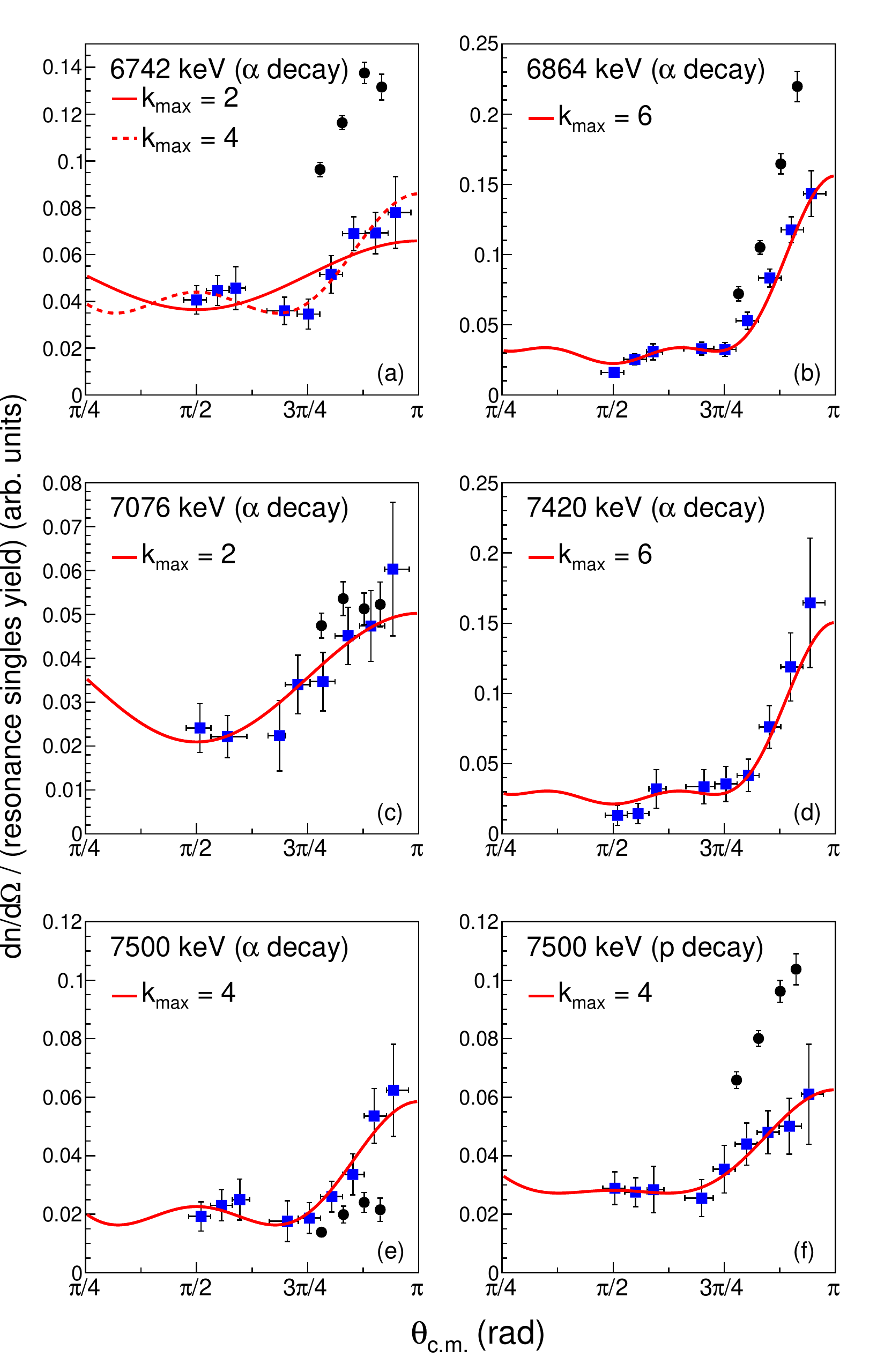}
    \caption{(Colour online) Triton-alpha and triton-proton angular correlation probabilities from the \fht($\alpha$)$^{15}$O and \fht(p)$^{18}$F reactions for the \nen\ states listed above. The squares (blue) are experimentally determined values with associated uncertainty. Error bars in $\theta_{\mathrm{c.m.}}$ represent the width of the angular bin. Even legendre polynomial terms are fitted to the data and the maximum order of the summation $k_{max}$ is indicated. The best fit is shown by the solid line (red). Circles (black) are from a similar experiment performed by Visser~\textit{et al.}~\cite{Visser2004}.}
    \label{fig:AngDist}
\end{figure}

The minimum order of Legendre polynomial needed to fit the data was decided on the goodness of fit achieved with progressively higher values of $k_{max}$ whilst maintaining a reasonable number of degrees of freedom. The best fits are shown in Fig.~\ref{fig:AngDist}, together with the data from Visser et al.~\cite{Visser2004} for comparison (see Section~\ref{sec:BR} for discussion). The theoretical fits provide an overall very good description of the experimental angular correlation. Each \nen\ state is now discussed individually.

The \nen\ state at $E_x=6742$~keV is the first above the proton threshold to be meaningfully analysed. This state was first observed in the $^{20}$Ne($^3$He,$^4$He)$^{19}$Ne reaction and its angular distribution indicates a $J^\pi=3/2^-$, (1/2$^-$) assignment~\cite{Garrett1970}. Based on mirror symmetry arguments the $J^\pi=3/2^-$ assignment was confirmed~\cite{Utku1998}. The maximum order of the summation in Eq.~\ref{eqn:AngDistSimp} for a state having $J^\pi=3/2^-$ is $k_{max}=2$, and the corresponding best fit ($\chi^2_\nu=13.8/7$) of the experimental data is represented by the solid line in Fig.~\ref{fig:AngDist}. Such value for the reduced $\chi^2$ corresponds to a $p$-value of 0.054 slightly greater than 0.05. This indicates that the present data is compatible with a $J^\pi=3/2^-$ assignment even though the angular correlation would be better described if one would consider $k_{max}=4$ (implying $J\geq5/2$) as shown by the dashed line curve in Fig.~\ref{fig:AngDist}.

The level at 6864~keV in \nen\ has been assigned a spin and parity $J^\pi=7/2^-$ based on mirror symmetry arguments~\cite{Utku1998}, which was further confirmed by the angular correlation analysis of Visser~et~al.~\cite{Visser2004}. The sum over the Legendre polynomials in Eq.~\ref{eqn:AngDistSimp} is limited to $k_{max}=6$ in case of a $J^\pi=7/2^-$ \nen\ state decaying in the $\alpha$-particle channel. The best fit ($\chi^2_\nu=4.7/5$) presented as a solid line in Fig.~\ref{fig:AngDist} shows a remarkably good description of the experimental data which confirms the spin and parity assignment $J^\pi=7/2^-$.

The 7076 keV state is one of the best studied resonance in the p+\fle\ system. It is known to have a spin parity of 3/2$^{+}$ with well measured partial and total widths~\cite{Bardayan2001}. The only possibility for the orbital angular momentum of the emitted $\alpha$-particle is $l = 1$, which implies $k_{max}=2$. The best fit ($\chi^2/\nu=0.7$) in these conditions is represented by the red solid line which supports an assignment of $J=3/2$ for the \nen\ state at 7076~keV. Unfortunately, proton decay from the state was only partially observed in the D1 and D2 DSSSDs preventing a comprehensive analysis of its distribution.

The 7420 keV state was observed in a proton resonant elastic scattering experiment, and the subsequent $\mathcal{R}$-matrix analysis found that it had most likely a spin and parity assignment $J^\pi=7/2^+$~\cite{Bardayan2004}. As in the case of the \nen\ state at $E_x=6864$~keV the angular correlation must be described with $k_{max}=6$. The best fit ($\chi^2_\nu=4.3/5$) shown in Fig.~\ref{fig:AngDist} as a solid red line compares very well the experimental angular correlation, thus supporting the spin and parity assignment $J^\pi=7/2^+$. Note that this level was not observed in another proton resonant elastic scattering and was concluded not to exist~\cite{Murphy2009}. However no other known \nen\ levels could constitute the observed peak in the current data. The spin and parity assignment obtained in this work being consistent with previous determination also adds support to its concluded existence from this work.

The 7500 keV state is strongly populated and it is the first \nen\ level with sufficient proton decay strength to allow the $t-p$ angular correlation to be extracted. This state was first observed with the $^{19}$F($^3$He,t)$^{19}$Ne reaction~\cite{Utku1998} and its $J^\pi=5/2^+$ assignment comes from pairing with a known $5/2^+$ state in \fln\ based on similar excitation energies~\cite{Nesaraja2007}. This spin and parity assignment was later confirmed by an $\mathcal{R}$-matrix analysis of proton resonant elastic scattering data~\cite{Murphy2009}, and in another coincidence measurement~\cite{Dalouzy2009}. For both the proton and $\alpha$-particle decay channels, the angular correlation is limited by $k_{max}=4$ for a $J^\pi=5/2^+$ emitting state. Best fits are represented as solid red lines in Fig.~\ref{fig:AngDist} and the excellent agreement with the experimental data supports a $J^\pi=5/2^+$ assignment.

\subsection{Branching ratios} \label{sec:BR}
For each experimental angular correlation analysed in Fig.~\ref{fig:AngDist} the associated $\alpha$-particle or proton branching ratio for the corresponding \nen\ state was obtained by integration of the theoretical correlation function over the full solid angle. We found that our branching ratios are consistently lower than previous values reported in the literature~\cite{Utku1998,Visser2004,Bardayan2004}. This observation is consistent with the present angular correlations being usually lower than those of Visser~et~al.~\cite{Visser2004} reported in Fig.~\ref{fig:AngDist}. The origin of this issue has been pinned down to an electronic problem affecting the coincidence event efficiency. This effect was found to be independent of both the focal plane position and the energy deposited in the silicon. The well studied \nen\ state at 7076~keV was therefore used as a benchmark for the branching ratio. Considering partial and total widths $\Gamma_p=15.2$~(1)~keV, $\Gamma_\alpha=23.8$~(12)~keV and $\Gamma=39.0$~(16)~keV~\cite{Bardayan2001, Bardayan2004}, the multiplication factor which must be applied to our data in order to reproduce the known $\alpha$-particle branching ratio ($\Gamma_\alpha/\Gamma=0.61$~(2)) for the \nen\ state at 7076~keV is $1.58\pm0.14$.

\begin{figure}[htpb]
	\includegraphics[width=\columnwidth]{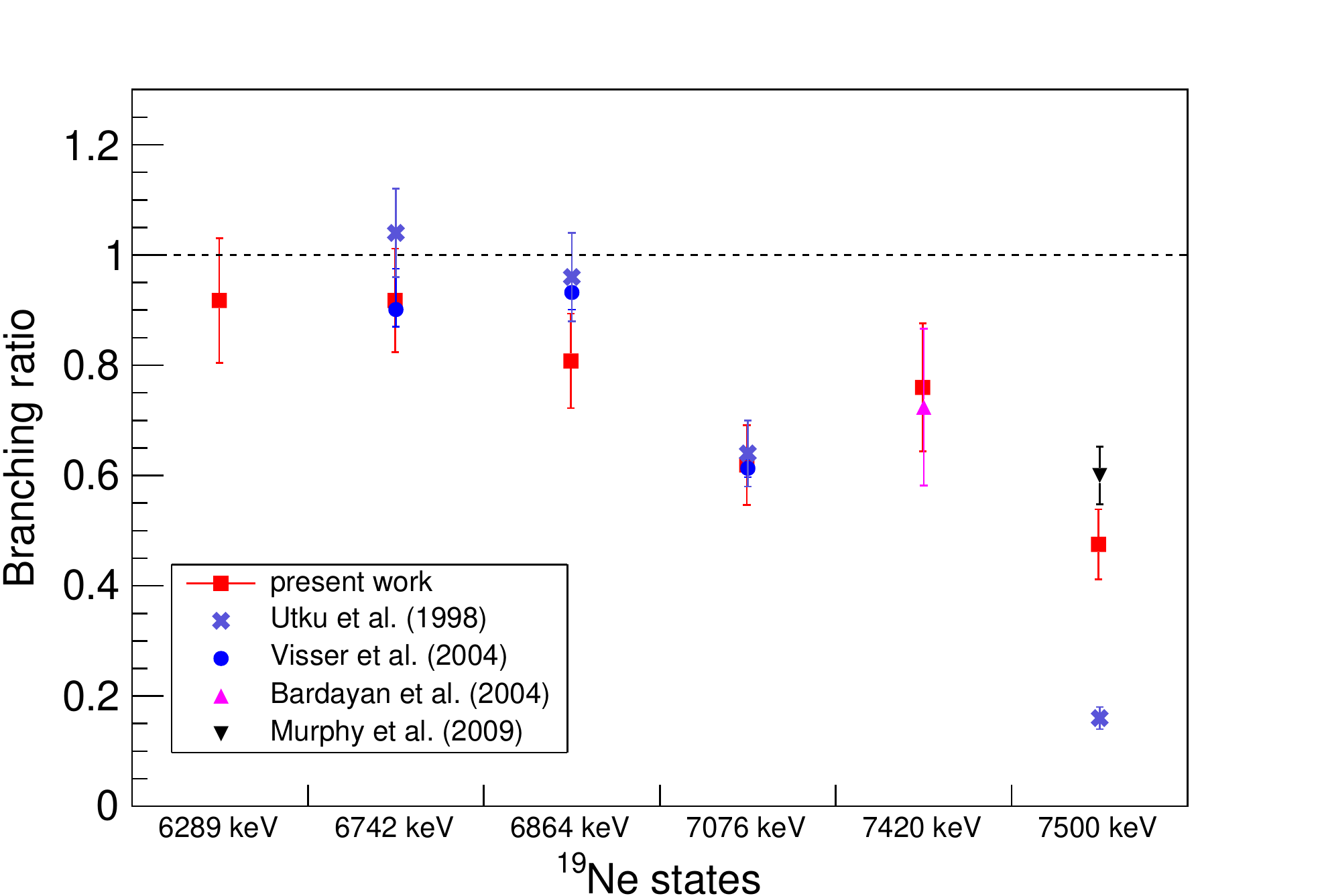}
	\caption{(Colour online) Alpha-particle decay branching ratios from the present work (red) are displayed with previous results from the literature for comparison.}
	\label{fig:BR}
\end{figure}

Branching ratios from the present work are reported in Tab.~\ref{tab:BR} together with results from previous works. The uncertainty associated to our branching ratio determination arises from the combined effect (quadratic sum) of the correction factor uncertainty and from the propagation of uncertainties of the angular correlation fit parameters when integrating over $4\pi$~sr. The comparison between different data sets is shown in Fig.~\ref{fig:BR} and good agreement is observed between our data and previous measurements. 

\begin{table*}[!htpb]
  \caption{\label{tab:BR}
     Alpha-particle branching ratios from the present work, and comparison with values reported in the literature. Excitation  energies and spin and parity assignment comes from literature unless otherwise stated. Resonance energies are given with respect to the p+\fle\ threshold ($S_p=6410.0$~(5)~keV).}
  \begin{ruledtabular}
  {\def\arraystretch{1.2}
  \begin{tabular}{ccccccccc}
    $E_x$ & $E_r^{cm}$ & $J^\pi$ & \multicolumn{5}{c}{$\Gamma_\alpha/\Gamma$ branching ratio} \\ \cline{1-3} \cline{4-8}
    (keV) &   (keV)    &         & Present & Utku et al.~\cite{Utku1998} & Visser et al.~\cite{Visser2004} & Bardayan et al.~\cite{Bardayan2004} & Murphy et al.~\cite{Murphy2009} \\ \hline
    6289\footnotemark[1]  & -121 & $>7/2$\footnotemark[2]  & 0.92 (11) &          & &           & \\
    6742  &  332 & $3/2^-$ & 0.92 (9)  & 1.04 (8) & $0.901^{+0.074}_{-0.031}$ &           & \\
    6864  &  454 & $7/2^-$ & 0.81 (9)  & 0.96 (8) & $0.932^{+0.028}_{-0.031}$ &           & \\
    7076  &  666 & $3/2^+$ & 0.62 (7)  & 0.64 (6) & 0.613 (16)  & 0.61 (2)  & \\
    7420  & 1010 & $7/2^+$ & 0.76 (12) &          & & 0.72 (14) & \\
    7500  & 1090 & $5/2^+$ & 0.47 (6)  & 0.16 (2) & &           & 0.60 (5) \\
  \end{tabular}    
  }
  \end{ruledtabular}
  \footnotetext[1] {This state is probably a doublet, see text for discussion.}
  \footnotetext[2] {From the present analysis.}
\end{table*}

The only exception is the \nen\ state at 7500~keV for which the present determination of the $\alpha$-particle branching ratio $\Gamma_\alpha/\Gamma=0.47~(6)$ is in agreement within $2\sigma$ with the measurement of Murphy et al.~\cite{Murphy2009} but disagrees with Utku et al.~\cite{Utku1998} who obtain $\Gamma_\alpha/\Gamma=0.16~(2)$. The proton branching ratio for the 7500-keV state could also be extracted from the corresponding angular correlation shown in Fig.~\ref{fig:BR} and we obtain $\Gamma_p/\Gamma=0.66$~(7), while Utku~et~al.~\cite{Utku1998} obtain $\Gamma_p/\Gamma=0.84~(4)$. The present proton and $\alpha$-particle branching ratios sum to 1.13~(9) which is compatible within two sigma with unity, and strengthens the reliability of the present analysis.

One possibility for this discrepancy could originate from the angular correlation analysis in Ref.~\cite{Utku1998}. The experimental angular correlations for the 7500~keV state (not shown in their paper) is restricted to a small angular range sampled by three detectors centered at laboratory angles of \ang{90}, \ang{110} and \ang{145}. The angular correlations are then independently fitted with a linear combination of the first three Legendre polynomials, thus implying three free parameters for three data points. The 7500~keV state is now known to have $J^\pi=5/2^+$~\cite{Murphy2009}, which implies $k_{max}=4$ and thus limits the sum in Eq.~\ref{eqn:AngDist} to the first three Legendre polynomials, confirming the number of free parameters used in Utku~et~al. analysis. However in their procedure Utku~et~al. don't consider the $\sum_{m}P(m)=1$ relation between the magnetic substate population which can lead to erroneous shape of the angular correlation function and biased determination of the branching ratios.
 
Another reason for the origin of the discrepancy with Utku et al. may be related to a possible contamination in the present data from the neighbouring state at 7531~keV. If one combines the individual branching ratio determined by Utku et al.~\cite{Utku1998} for the two \nen\ states at 7500- and 7531-keV with their relative population as observed (FWHM = 24 keV) in their Fig.~1, one would get for these two states combined branching ratios of $\Gamma_\alpha/\Gamma=0.27(3)$ and $\Gamma_p/\Gamma=0.73(4)$, in much better agreement with the results from the current work. A similar effect may also affect the data of Murphy~et~al~\cite{Murphy2009} where the energy resolution does not allow to separate both states.


\subsection{The sub-threshold state at $E_x$~=~6289~keV}
\begin{figure}[htpb]
	\includegraphics[width=\columnwidth]{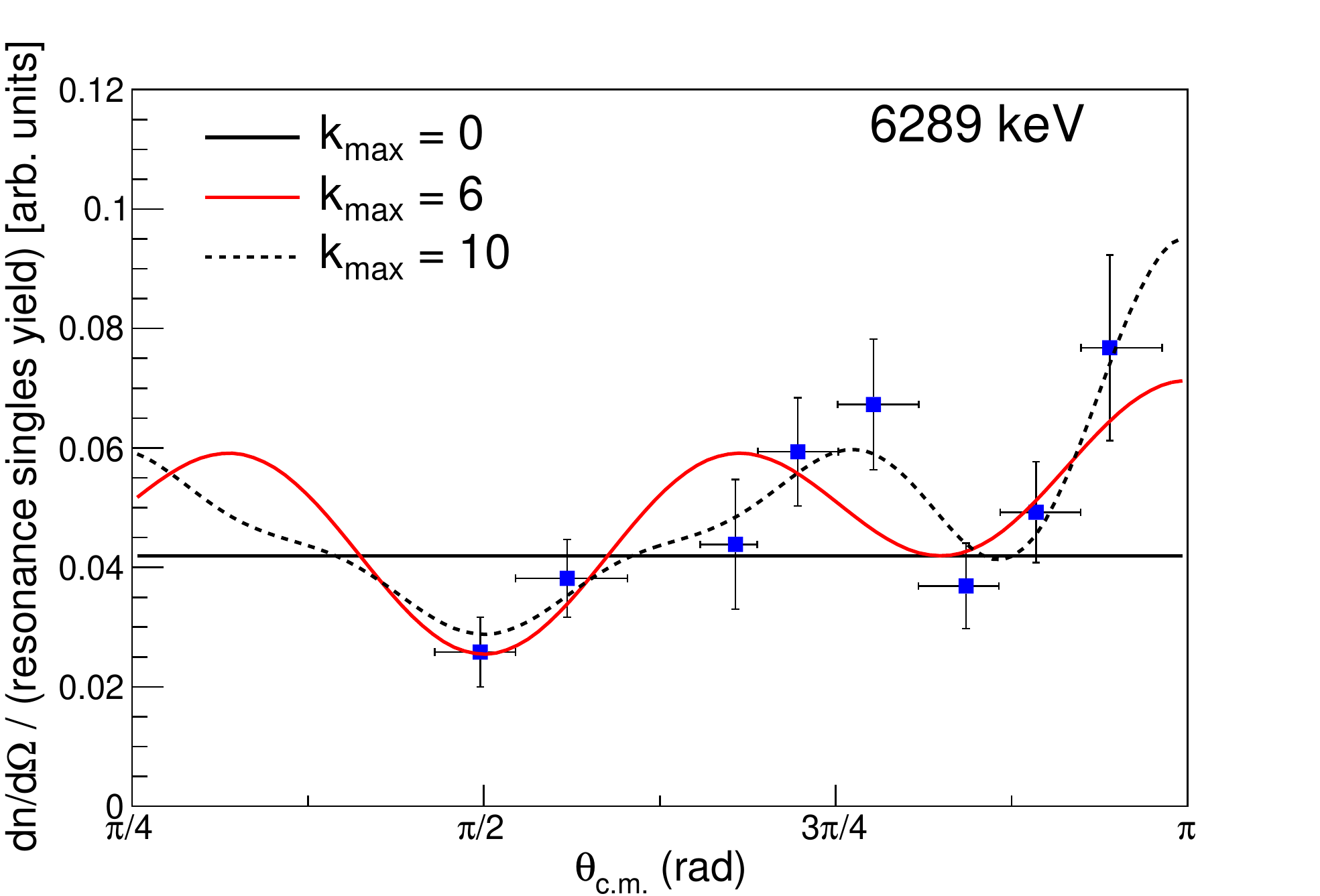}
   \caption{(Colour online) Triton-alpha angular correlation from the decay of the \nen\ doublet at $E_x=6289$~keV. The blue squares are experimentally determined values with associated uncertainty. Error bars in $\theta_{\mathrm{c.m.}}$ represent the width of the angular bin. Best fits of the angular correlation function are represented for three different values of $k_{max}$ (see text).}
	\label{fig:6289}
\end{figure}

The angular correlation of the \nen\ 6289~keV state is represented in Fig.~\ref{fig:6289}. While there are evidences of a close doublet at this energy~\cite{Parikh2015,Hall19} separated by about 12~keV~\cite{Parikh2015}, the energy resolution and the line shape asymmetry of the present data does not allow to separate them. Then the angular correlation in Fig~\ref{fig:6289} embeds the two possible contributions and a combined analysis is performed. The minimum value of $k_{max}$ providing a good description of the angular correlation was determined using $\chi^2_\nu$ $p$-value hypothesis testing. The null hypothesis was first chosen to correspond to an isotropic correlation ($k_{max}=0$) and was accepted if the $\chi^2_\nu$ $p$-value was greater than 0.05. For a smaller $p$-value the null hypothesis (isotropic angular correlation) was rejected and $k_{max}=2$ was considered as the new null hypothesis. This procedure was repeated until $p>0.05$ and the corresponding $k_{max}$ was considered as the minimum value providing a good description of the data. The isotropic case ($k_{max}=0$), which would imply $J=1/2$, is clearly rejected as can be observed in Fig~\ref{fig:6289} (solid black line) with $p=1.5\times10^{-3}$. The first case compatible with the experimental angular correlation is obtained for $k_{max}=6$ (red solid line) with $p=0.16$. According to the definition of $k_{max}$ a value of 6 corresponds to a spin of 7/2. We therefore conclude from the current analysis that the peak corresponding to excitation energies about 6289~keV behaves as a state with a rather high spin $J\geq7/2$. The calculation of the correlation function with $k_{max}=10$, corresponding to an initial spin $J=11/2$, is also shown in Fig.~\ref{fig:6289} for comparison (dashed black line). As expected the angular correlation is better reproduced ($p=0.39$) since additional free parameters are considered in the fitting procedure. As for the other \nen\ states, the $\alpha$-particle branching ratio is calculated with the same procedure and yields $\Gamma_\alpha/\Gamma = 0.92\pm0.11$ for this $E_x=6289$~keV doublet. As expected the branching ratio is compatible with 1 since at this energy only the $\alpha$-particle decay channel is open.

As suggested~\cite{Parikh2015, Hall19}, the existence of a second state close in energy is readily explained by all measurements of the state performed to date. Data by Adekola \textit{et al.}~\cite{Adekola2011a} and Bardayan \textit{et al.}~\cite{Bardayan2015} were taken using $^{18}$F(d,n)\nen\ and $^{20}$Ne(p,d)\nen\ respectively. Data in this work and that of Laird \textit{et al.}~\cite{Laird2013} as well as Kahl \textit{et al.}~\cite{Kahl19}, populated $^{19}$Ne through \fht. It is expected that different reaction mechanisms may preferentially populate different states depending on the required $l$-transfer. Two possibilities remain, therefore, for interpreting the angular distributions in this work. Either the J$^{\pi}$ = 1/2$^{+}$ is far lower in intensity and the distributions represent the spin from the second state, or both are populated to a non-negligible proportion and the $\alpha$-particle decay measurements are mixing from both. Unfortunately, given the resolution and asymmetry of the focal plane, resolving two peaks at B$\rho$ = 0.975 Tm with the predicted 12 keV difference was not possible. The analysis from this work can confirm, however, that the observed resonance cannot be a single state of J$^{\pi}$ = 1/2$^{+}$.

\section{Astrophysical $S$-factor} \label{sec:Rate}
An analysis of the \fpa\ reaction at novae temperatures was conducted using the $\mathcal{R}$-matrix formalism~\cite{Lane1958} with the AZURE2 code~\cite{Azuma2010}. While the focus is on the impact of sub-threshold states at 6.008-, 6.132- and 6.286~MeV, the contribution of influential resonances in the Gamow window is also calculated. This includes the 3/2$^+$ states just above the p+\fle\ threshold~\cite{Hall19}, the $3/2^-$ state at 6.740~MeV~\cite{Bardayan2002}, the $3/2^+$ state at 7.075~MeV~\cite{Bardayan2001} and the $1/2^+$ state at 7.879~MeV~\cite{Adekola2012}.

\subsection{The doublet at $E_x$~=~6.29~MeV}
Taking together all the experimental data available, it is clear that in the region of 6.29 MeV two states are present, one of high spin, the other low spin. It is assumed that the high spin component is 11/2$^+$ as reported by Hall et al.~\cite{Hall19}, this assignment being supported by the present analysis. However, a high spin component will not contribute to the astrophysical reaction rate and so a firm assignment is not required. For the low spin state, we prefer a spin assignment of 1/2 over 3/2, based on the clear signature from the $^{20}$Ne(p,d)$^{19}$Ne study~\cite{Bardayan2015}. Although Kahl et al.~\cite{Kahl19} prefer a 3/2 assignment based on the required mirror energy difference, we find this argument less compelling given the large shifts already observed, e.g. the 208 keV between the 11/2$^+$ states at 6.292 and 6.500 MeV in $^{19}$Ne and $^{19}$F, respectively. Furthermore Dufour and Descouvemont~\cite{Dufour2007} found large differences to be possible for $s$-wave states (i.e. 1/2$^+$ or 3/2$^+$) with large spectroscopic factors. Alpha-particle widths and Asymptotic Normalization Coefficients (ANC) for the 1/2$^+$ states is taken from Ref.~\cite{Kahl19}. 

\subsection{The level at $E_x$~=~6.132~MeV}
This state has been populated by the \fht\ reaction~\cite{Utku1998,Laird2013,Parikh2015} and its angular distribution found to be indicative of a $(3/2^+)$ or $(5/2^-)$ state~\cite{Laird2013}. A $\Delta L=0$ transition was observed at 6130~(5)~keV by Kahl et al. and a $J^\pi=1/2^+$ assignment favored, though $J^\pi=3/2^+$ is not discarded~\cite{Kahl19}. Furthermore, the analysis of the $p-\alpha$ angular correlation of this state populated through the $^{19}$Ne(p,p$^\prime$)$^{19}$Ne($\alpha$)$^{15}$O reaction favors a $J=3/2$ assignment~\cite{Dalouzy08}. All observations can, therefore, be reconciled if this level is a $J^\pi=3/2^+$ state, and we have used this assignment in the $\mathcal{R}$-matrix calculations. This assignment, however, implies that either there is an, as yet, unidentified 3/2$^+$ state in $^{19}$F, or one of the two 3/2$^+$ states suggested by Hall et al. is mis-assigned. Alpha-particle widths and ANC for this state is taken from Ref.~\cite{Kahl19}. 

\subsection{The broad state at $E_x$~=~6.008~MeV}
Based on the extracted width, the spin-parity is $J^\pi=1/2^-, 1/2^+$ or $3/2^+$, and we consider possible analogue states in $^{19}$F for each case here. In the case of a $J^\pi=1/2^-$ assignment the only possibility for a mirror connection with a known state would be with the 6.429~MeV state in \fln. This connection would require a rather large, but not prohibitively so, mirror energy difference (more than 400~keV). There is some evidence, however, that the 6.429~MeV state is paired with the 6.439~MeV state in \nen\ based on the work of Utku et al.~\cite{Utku1998}.

If the broad state has a $J^\pi=1/2^+$ assignment, this raises the question of whether it can be associated to the broad 1/2$^+$ state at 6.001~MeV ($\Gamma=231$~keV) predicted using the Generator Coordinate Method (GCM)~\cite{Dufour2007}. The energy and total width of the predicted state rely on the association of the theoretical GCM 1/2$^+$ state in \fln\ with the experimentally known 1/2$^+$ state at 5.94~MeV. It should, in fact, be associated to the known $\alpha$-cluster state at 5.34~MeV ($\theta^2_\alpha=0.53$)~\cite{Descouvemont87}, which then modifies the parameters predicted by Dufour et al.~\cite{Dufour2007} such that the state is not now expected to be broad. Experimentally, there is strong evidence that in this energy region the \fln\ state at 5.34~MeV has a much stronger $\alpha$-particle clusterization than the 5.94~MeV state. First, the $\alpha$-particle width for the 5.34~MeV state has been determined experimentally ($\Gamma_\alpha=1.3$~(5)~keV) and compared to single-particle width calculated with a potential model, leading to $\theta^2_\alpha\approx0.4$~\cite{Wilmes02}, in agreement with theoretical predictions~\cite{Descouvemont87}. Other experimental work finds $\Gamma_\alpha=2.51$~(10)~keV~\cite{DiLeva17} also supporting a strong $\alpha$-cluster contribution for the 5.34~MeV state. Second, the $^{15}$N($^7$Li,t)$^{19}$F reaction was studied at bombarding energies of 15~MeV and 20~MeV~\cite{Middleton70}. In both cases the 5.34~MeV state is very well populated, and  while the 5.94~MeV is not labeled (see Figs.~7 and 10 in Ref.~\cite{Middleton70}) it can be estimated from these energy spectra that its $\alpha$-particle spectroscopic factor is at least three times lower than for the 5.34~MeV. This indicates that the 5.94~MeV state has a much smaller $\alpha$-cluster configuration than the 5.34~MeV state. The large $\theta^2_\alpha$ deduced for the broad state under consideration (see Sec.~\ref{sec:Broad}) could be an indication for being the mirror of the 5.34~MeV in \fln, however this would require a very large energy shift of 1.15~MeV with respect to their respective $\alpha$-particle threshold. Even though large energy shift are possible for strongly clusterized $s$-wave states~\cite{Dufour2007}, a shift of 1.15~MeV would be surprising and the above mentioned analog pairing is very unlikely.

Finally, considering the case of a $J^\pi=3/2^+$ assignment a counterpart should have been predicted by the GCM since its experimental dimensionless reduced width is large. Indeed the GCM predicts a 3/2$^+$ state in \fln\ which is associated to the experimentally known 3/2$^+$ at 5.501~MeV~\cite{Dufour2007}. However there is some evidence that this state is the analog of the 5.463~MeV state in \nen~\cite{Kahl19}.

Given the lack of knowledge of the spin and parity assignment of the tentative broad state at $E_x=6.008$~MeV, all three possible assignments have been considered in order to evaluate its potential contribution above the p+\fle\ threshold. Its reduced proton width can be calculated using the following relation~\cite{Ili08} $\gamma^2_p = \theta^2_p \times \gamma^2_{p,Wigner}$, where $\gamma^2_{p,Wigner}= 3\hbar^2/(2\mu r^2)= 2.246$~MeV is the Wigner limit for the proton reduced width evaluated at a channel radius $a_p=5.4$~fm. A dimensionless proton reduced width $\theta^2_p=1.8\times10^{-3}$ is considered using the results from the systematic study of Ref.~\cite{Pog13}. This leads to ANC values of 1.19~fm$^{-1/2}$ and 0.64~fm$^{-1/2}$ in case of a proton in the $s$- or $p$-shell, respectively.

The proton transfer reaction $^{18}$F(d,n)$^{19}$Ne($\alpha$)$^{15}$O~\cite{Adekola2011} can also be used to assess the potential importance of the $E_x=6.008$~MeV state. In that work there is no evidence of any significant proton strength between $E_x=5.49$ and 6.09~MeV (see Fig. 5(b)). We roughly estimate that a number of coincident events greater than $\approx30$ for the contribution of the $E_x=6.008$~MeV state would have been detected. Assuming a similar angular distribution for the $E_x=6.008$~MeV state as the known $s$-wave state at $E_x=6.289$~MeV, a spectroscopic factor $(2J+1)S_p \lesssim 0.03$ is deduced for the $E_x=6.008$~MeV state. The single-particle proton ANC for this state was calculated assuming a Woods-Saxon potential well with geometry $(r,a)$ = (4.5 fm, 0.53 fm) and this gave ANC$_{s.p.} = 17.1$~fm$^{-1/2}$. This leads to ANC values of  1.47~fm$^{-1/2}$ and 2.08~fm$^{-1/2}$ for a $J^\pi = 3/2^+$ and 1/2$^+$ assignment, respectively.

\subsection{Results and discussion}
\begin{figure}[!htpb]
  \includegraphics[width=0.5\textwidth]{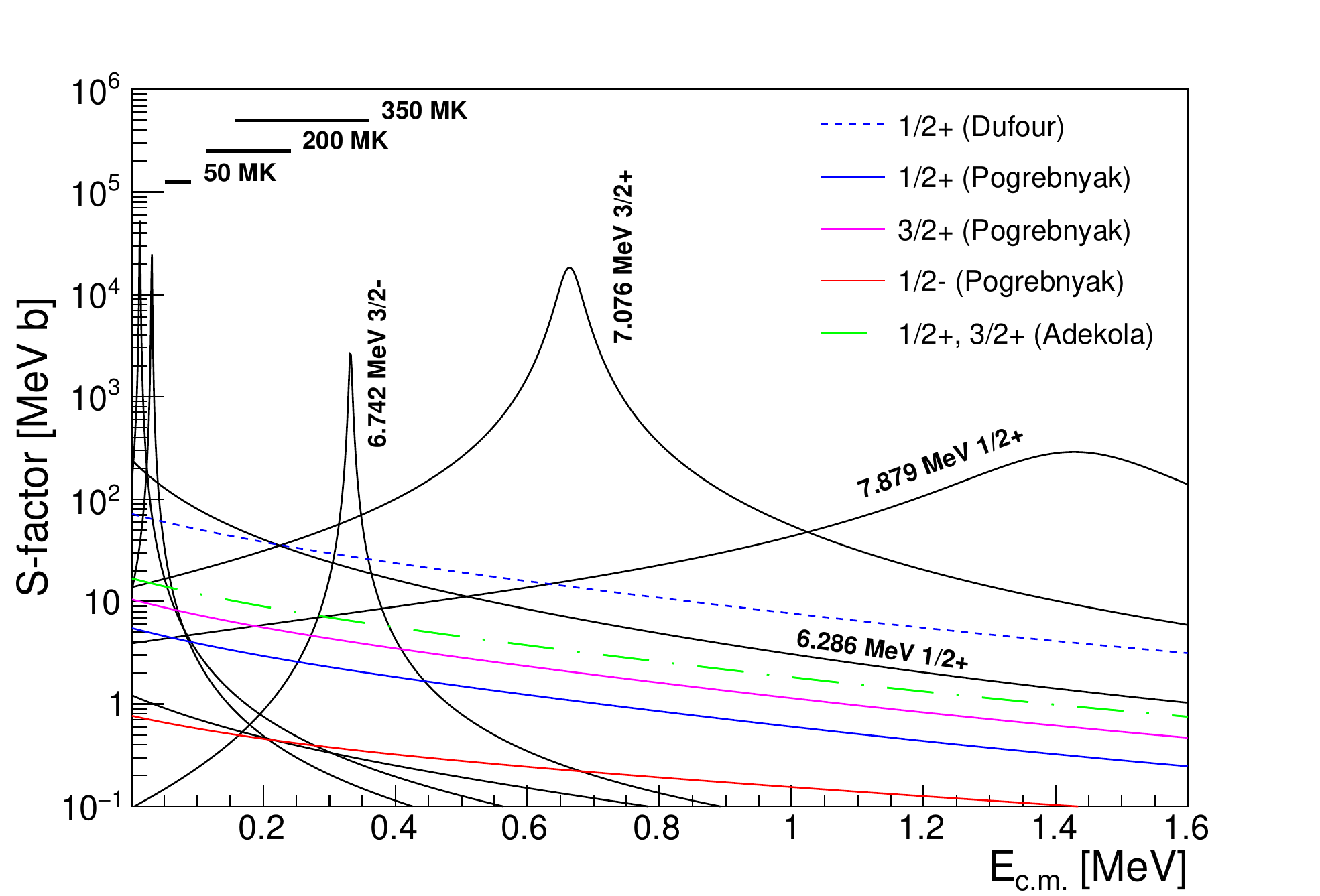}
  \caption{\label{fig:subtrhreshold_6MeV}
  (Colour online) Astrophysical $S$-factor for the \fpa\ reaction as a function of the center-of-mass energy. $\mathcal{R}$-matrix calculations for the 6.008~MeV state are represented in color lines for different spin-parity assumptions. Calculations with three different ways of estimating the proton ANC values are presented in case of a $1/2^+$ assignment (see labels in the legend and text for discussion). The contribution of the most influential resonances is represented in solid black lines for comparison purposes.}
\end{figure}

$\mathcal{R}$-matrix calculations using channel radius $a_p=5.4$~fm (entrance) and $a_\alpha=6.1$~fm (exit) are presented in Fig.~\ref{fig:subtrhreshold_6MeV} for the above mentioned \nen\ states with the exception of the 6.132~MeV state whose contribution is lower than 1 MeVb for all center of mass energies. The dash-dotted line represent the expected contribution for the sub-threshold state at 6.008~MeV with the ANC values computed using the experimental constraints from the work of Ref.~\cite{Adekola2011}, while the solid lines are estimates based on the systematic study of Ref.~\cite{Pog13}. Given the magnitude of the $S$-factor it is very unlikely that this state has a strong impact in the \fpa\ reaction rates since the contribution of other resonances (solid black lines) in the Gamow window dominate. It is worth noting that the proton ANC values used for the estimate of the 6.008~MeV state based on the systematic study of Ref.~\cite{Pog13} can be uncertain by large factors. This is related to the scatter of the dimensionless proton widths reported in that study, and this could have significant impact on the potential role of the 6.008~MeV state. 

Although it is not clear whether or not the 6.008~MeV state can be associated to the broad \nen\ state predicted theoretically using the GCM~\cite{Dufour2007}, the consequences of such a possibility can still be explored. A theoretical partial width $\gamma^2_p = 1.95\times10^{-3}$ is reported for a channel radius $a_p = 10.1$~fm~\cite{Dufour2007} which corresponds to an ANC of 4.3~fm$^{-1/2}$. The contribution to the $S$-factor of the 6.008~MeV state using this ANC value has been calculated and is shown in Fig.~\ref{fig:subtrhreshold_6MeV} by the dashed blue curve. As expected the contribution of the 6.008~MeV state to the \fpa\ reaction rate could be much more important with these parameters.

The broad state at $E_x=6.008$~MeV will not play a role unless it has a large proton reduced width, similar to that of the GCM prediction. Such a value is not necessarily in contradiction to the systematic study of Pogrebnyak et al.~\cite{Pog13} given the large scatter in those data and the trend of increasing reduced width for lower mass number. Similarly, the GCM predictions are not necessarily in contradiction with the estimate based on the Adekola et al. work~\cite{Adekola2011} given the rater crude estimate of the maximum contribution of the $E_x=6.008$~MeV state given here.

\section{Conclusion} \label{sec:Conc}
The level scheme of \nen\  has been studied through the coincident detection of tritons and $\alpha$-particles from the $^{19}$F($^{3}$He,t)$^{19}$Ne$^*$($\alpha$) reaction. The results support the presence of a doublet at around 6.29 MeV consisting of a high spin (likely 11/2$^+$) state and a low spin (1/2$^+$) state. The state at 6.130 MeV was observed but due to the experimental resolution, the angular correlation could not be separated from the much stronger 6.014 MeV state. The 6.130 MeV has been assumed to be 3/2$^+$ but experimental confirmation of this assignment is needed. Evidence for a broad state at 6.01 MeV was found. Due to the large observed width, this state is likely to be low spin.  Branching ratios were determined for states between 6.289 and 7.5 MeV, and are in good agreement with the literature. 

 $\mathcal{R}$-matrix calculations have been performed 
 showing the contribution of key states in \nen\ compared to estimates of the possible contribution of the tentative broad ($\Gamma$ = 124 keV) state at 6.008 MeV. These calculations indicate that, if it is to be significant for the \fpa\ reaction rate, the state must be 1/2+ or 3/2+ and its reduced proton width of similar magnitude to that predicted by Dufour and Descouvemont. Given the tentative evidence presented here, a high-resolution, high-statistics measurement is needed to provide clarification on the origin of the excess of counts observed in this region.
 
 It is clear that significant gaps in our knowledge of the level scheme of \nen\  and indeed $^{19}$F remain, below the proton threshold as well as above. The connection of analogue states above 5 MeV is far from complete and the data suggest there may be unobserved and/or mis-assigned states in both nuclei. Not all of the missing information is required to constrain the \fpa\ reaction rate however. The important parameters remain the proton widths of the 3/2$^+$ states above the proton threshold and the interference terms between $l=0$  states with the same spins.  However, as the interference terms can only be determined by a direct measurement, higher statistics measurements below the 331 keV resonance are most critical to reducing the uncertainty on the \fpa\ reaction rate at nova temperatures.

\begin{acknowledgments}
The continued support of the staff of the Tandem-Alto facility is gratefully acknowledged. NdS and AML acknowledge very fruitful discussions with Pierre Descouvemont about the properties of sub-threshold \nen\ states, and with Michael Bentley on mirror energy differences. AML acknowledges the support of the Science and Technology Facilities Council and the Royal Society. This article is based upon work from the “ChETEC” COST Action (CA16117), supported by COST (European Cooperation in Science and Technology).
\end{acknowledgments}

\bibliographystyle{apsrev4-1}
\bibliography{f19he3t_v2}

\end{document}